\documentclass[prb,twocolumn,showpacs,citeautoscript,amsmath,amssymb,superscriptaddress]{revtex4}
\usepackage{graphicx}
\usepackage{dcolumn}
\usepackage{bm}
\usepackage{amsmath}
\usepackage{amssymb}
\usepackage{epsfig}
\usepackage{color}

\begin{document}
\title{Influence of interface transmissivity and inelastic scattering on the electronic entropy and specific heat of diffusive SNS Josephson junctions}

\author{Hassan Rabani}
\email{h_rabani@sci.ui.ac.ir}
\affiliation{Department of Physics, Faculty of Sciences, University of Isfahan, 81744 Isfahan, Iran}
\affiliation{NEST CNR-INFM and Scuola Normale Superiore, I-56126 Pisa, Italy}
\author{Fabio Taddei}
\affiliation{NEST CNR-INFM and Scuola Normale Superiore, I-56126 Pisa, Italy}
\author{Francesco Giazotto}
\email{giazotto@sns.it}
\affiliation{NEST CNR-INFM and Scuola Normale Superiore, I-56126 Pisa, Italy}
\author{Rosario Fazio}
\affiliation{NEST CNR-INFM and Scuola Normale Superiore, I-56126 Pisa, Italy}

\begin{abstract}
We study theoretically the electronic entropy and specific heat in diffusive superconductor-normal metal-superconductor 
(SNS) Josephson junctions. In particular, we consider the influence of non-idealities occurring in an actual experiment, such as the presence of barriers at the NS interfaces, the spin-flip and inelastic scattering in the N region and quasiparticle subgap states in the superconductors. We find that spin-flip and inelastic scattering do not have, for typical parameters values, a large effect. On the contrary, the presence of barriers suppresses the superconducting correlations in the N region, with the consequence that the entropy and the specific heat get reduced eventually to those in the absence of superconductivity for opaque interfaces. 
Finally we suggest an experiment and check that it is possible, under realistic conditions, to measure the dependence of electronic specific heat on the phase difference between the superconductors.
\end{abstract}
\maketitle

\section{Introduction}
\label{s1}
When a superconductor (S) is put in contact to a normal metal (N), Cooper pairs can leak across the 
interface. As a result there exists a non-vanishing pair amplitude in the normal metal. This is 
the so called {\em proximity effect} at the heart of the properties of SN hybrid systems \cite{degennes}. 
Already studied 
during the early days of superconductivity, the interest in proximity devices has been renewed when it 
became possible to study hybrid systems with dimensions comparable or smaller than the coherence length 
in the normal metal. Over the last years an intense theoretical and experimental activity has unveiled 
numerous interesting aspects of proximity-based devices (see Refs.~\onlinecite{editorHekking,Beenakker95,
Lambert96,Belzig,Taddei,ds}).
So far most of the attention has been devoted to their transport properties, although their 
thermal properties are of great importance especially in relation to the use of hybrid devices for electron 
cooling~\cite{Giazotto}.
Thermodynamic properties have been considered in Ref.~\onlinecite{Rabani} for a SNS structure showing that proximity effect may lead to a substantial deviation of
the specific heat from that in the normal state and that it can be largely tuned in magnitude by changing
the phase difference between the superconductors.
To confirm these predictions a measurement setup, which makes use of a highly-sensitive 
recently developed method~\cite{Bourgeois,Ong}, has been also suggested.

In the present paper we further elaborate on this problem.
Namely, we extend the results reported in Ref.~\onlinecite{Rabani} in order to investigate the effect of non-idealities occurring in an actual experiment, such as inelastic and spin-flip scattering, and the role of the barriers and the SN interfaces.
Moreover, as most probably the experiment will be performed embedding the SNS junction within superconducting loops, the signal will also collect the contribution due to the superconducting region, as well as the unavoidable contribution due to phonons.
We include such additional contributions in order to find the conditions under which they do not hinder the signature of proximity effect.

The paper is organized as follows. In Sec.~\ref{theo} we describe the general formalism based on the 
quasiclassical theory for diffusive SNS junctions. 
In Sec.~\ref{res} we present our results. We show the entropy and specific heat of the proximized N region in a SNS junction calculated as functions of temperature and superconducting phase difference, and compare with the ones relative to a metal in the normal and superconducting state.
Furthermore we explore the dependence of the entropy and specific heat on the strength of spin-flip and inelastic scattering, interface resistance and presence of subgap quasiparticle states.
Moreover, we check that the proposed experimental setup allows to measure the phase-dependent contribution to the specific heat of the SNS junction.
We summarize our main conclusions in Sec.~\ref{conc}.

\section{Theoretical Framework}
\label{theo}
\begin{figure}[t]
\includegraphics[width=0.35\textwidth,clip]{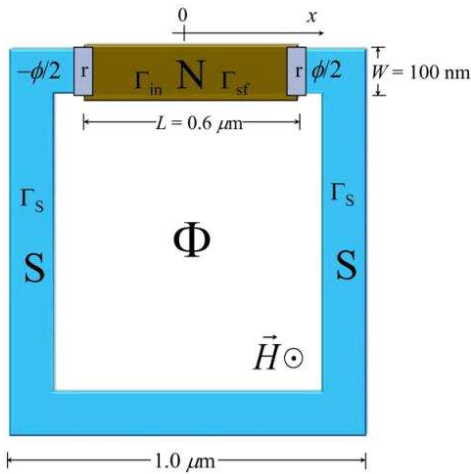}
\caption{(Color online) Schematic of a SNS junction embedded in a superconducting loop. The normal (N) metal region, of length $L$ and width $W$, is placed between two barriers  and subject to inelastic and spin-flip scattering whose rates are, respectively, $\Gamma_{\rm in}$ and $\Gamma_{\rm sf}$. $\Gamma_{\rm S}$ is the inelastic scattering rate in the superconductor (S). The perpendicular magnetic field $\vec{H}$ and the geometry allow a superconducting phase difference $\phi$ to appear between the S electrodes across the N region. $\Phi$ is the total flux threading the loop area and $x=0$ denotes the middle of the junction. The length values indicated are typical for a system made of a Cu as N and Al as S materials.}
\label{fig1}
\end{figure}

In this section we will introduce the theoretical formalism needed to calculate the entropy and the specific heat of the system depicted in Fig.~ \ref{fig1}, which consists of a superconducting loop interrupted by a SNS junction.
The N metal length is equal to $L$, and the phase difference $\phi$ across such junction is controlled through a perpendicular magnetic field $\vec{H}$.

The formalism of quasiclassical Green's functions has been proved to be a powerful tool to describe
mesoscopic superconductivity~\cite{Belzig}.
In the dirty limit the mean free path is much smaller than coherence length and the Eilenberger equations can be simplified to the Usadel equations~\cite{Usadel} which we use in this paper.
Although much simpler than the original microscopic version, 
the equations for quasiclassical Green's functions are still hard to solve analytically and numerical methods 
are usually unavoidable. 
The following two different parametrizations for the quasiclassical equations are often used~\cite{Belzig}.
The first one, and most widely employed, is the $\theta$ parametrization which, however, turns out to yield unstable differential equations
in certain cases (for example when the phase difference is close to $\pi$).
The second one is the Riccati parametrization which is particularly suited for 
time-dependent problems, and yields stable differential equations.
In the next subsections we will briefly introduce these parametrizations, and define the electronic entropy and the specific heat of the system.

\subsection{$\theta$ parametrization}
In the $\theta$ parametrization~\cite{Belzig}
the Usadel equations can be written as
\begin{eqnarray}
&\hbar D\partial^2_x\theta =-2i(E+i\Gamma_{\rm in})\sinh(\theta) +\frac{\hbar D}{2}
\left(\partial_x\chi \right)^2\sinh(2\theta) &\nonumber\\
&\text{sinh}(2\theta)\partial_x \theta\partial_x \chi+\text{sinh}^2(\theta)\partial_x^2\chi=0 , &
\label{retard}
\end{eqnarray}
where $\theta$ and $\chi$ are complex scalar functions of energy $E$ and position $x$,
$D$ is the diffusion coefficient and $\Gamma_{\rm in}$ describes a small inelastic scattering rate in the N region.
The energy $E$ is measured from the chemical potential of the superconductors.
The boundary conditions 
at the NS contacts for perfectly transmitting interfaces are~\cite{Laakso}
\begin{equation}
  \begin{array}[c]{rclcl}
    \theta(\pm L/2)&=&\theta_{\rm S}=\hbox{arctanh}[\Delta/(E+i\Gamma_{\rm S})]\;,\\
    \chi(\pm L/2)&=&\pm\phi/2\;,
  \end{array}\label{bc}
\end{equation}
where $L$ is length of the normal region, $\Delta$ is the temperature-dependent superconducting
order parameter, and $\Gamma_{\rm S}$ takes phenomenologically into account a possible smearing of the density of states (DOS) due to inelastic scattering in the superconductor~\cite{Laakso}. 
The latter might be due either to impurities or to inverse proximity effect induced by the nearby-contacted metal~\cite{pekola}.
We recall that $\phi$ denotes the phase difference between the S electrodes.
For the purpose of the present paper, the self-consistent solution of the 
order parameter is not enforced so that $\Delta$ is assumed constant in S and zero in the N region.
Furthermore, 
we assume the usual BCS temperature dependence for $\Delta$ with critical temperature 
$T_{\text{C}}=\Delta_0/(1.764k_{\text{B}})$, where $\Delta_0$ is the zero-temperature order parameter, 
and $k_{\text{B}}$ is the Boltzmann constant.
If $R_{\rm N}$ is the resistance of the N region coupled via a barrier with 
resistance $R_{\rm B}$ to a superconductor, the boundary condition for $\theta$, derived by 
Kupriyanov and Lukichev~\cite{Kupriyanov}, becomes
\begin{equation}
rL\,\partial_x\theta|_{x=\pm L/2}=\pm\sinh[\theta(\pm L/2)-\theta_{\rm S}]\; .
\label{rbc}
\end{equation}  
where $r=R_{\rm B}/R_{\rm N}$, and $\theta_{\rm S}$ is defined in 
Eq. (\ref{bc}).
The local DOS in the normal region can be obtained by
\begin{equation}\label{density}
   \mathcal{N}(x,E,T,\phi)=\mathcal{N}_{\text{F}} \mbox{Re}\left\{\cosh\left[\theta(x,E,T,\phi)\right]\right\},
\end{equation}
where $\mathcal{N}_{\text{F}}$ is the DOS per spin at the Fermi level in the absence of 
superconductivity.

\subsection{Riccati parametrization}
\label{ricc}
The Riccati parametrization \cite{Schopohl,Eschrig} has been employed by Hammer \emph{et al.}~\cite{Hammer} 
in order to calculate the supercurrent as well as the DOS for SNS junctions in the presence of opaque NS interfaces 
and spin-flip scattering. Here we just present the main formulas for completeness. 
The Usadel equations read
\begin{eqnarray}
\hbar D\partial^2_x \alpha_{1,2} + \hbar D(\partial_x \alpha_{1,2})^2 {\cal F}_{1,2}
\mp 2 \Gamma_{\rm sf} \alpha_{1,2}{\cal G}_{1,2}\nonumber\\
+2i (E+i\Gamma_{\rm in})\alpha_{1,2}=0,
\label{g-eq}
\end{eqnarray}
where
\begin{equation}
{\cal G}_{1,2}=\pm \frac{1-\alpha_{1} \alpha_{2}}{1+\alpha_{1}\alpha_{2}}\;; \hspace{1cm}
{\cal F}_{1,2}=-\frac{2\alpha_{2,1}}{1+\alpha_{1} \alpha_{2} }\;,
\label{GF}
\end{equation}
and $\alpha_1$ and $\alpha_2$ are complex scalar functions of energy $E$ and position $x$.
Here we have also introduced an effective spin-flip scattering rate $\Gamma_{\rm sf}$ which accounts for pair-breaking mechanisms, such as due to magnetic impurities or to the presence of a perpendicular magnetic field $H$.
For a thin normal region with small width $W$ (smaller than or comparable to the coherence length in the N 
metal $\xi=\sqrt{\hbar D/\Delta}$), such rate is given by $\Gamma_{\rm sf}= D e^2 H^2 W^2/(6 \hbar)$~\cite{Belzig2,Hammer}.
The boundary conditions for Eqs.~(\ref{g-eq}), which can include the transmission coefficient of the interface 
and the ratio $r$, are given in Ref.~\onlinecite{Hammer}.
Finally, the local DOS is given by
\begin{equation}
\mathcal{N}(x,E,T,\phi)=\mathcal{N}_{\text{F}} {\rm Re} \left[ {\cal G}_{1}(x,E,T,\phi)  \right].
\label{ricDOS}
\end{equation}

\subsection{Electronic entropy and specific heat}
The electronic specific heat ($C$) of the proximized N region of the SNS junction in Fig.~\ref{fig1} can be calculated through the expression
\begin{equation}
 C(T,\phi)=T\frac{\partial \mathcal{S}(T,\phi)}{\partial T}, 
 \label{Cv}
\end{equation}
where the electronic entropy ${\cal S}$ is defined as 
\begin{eqnarray}
\label{Sss}
\mathcal{S}(T,\phi)=-\frac{4k_{\rm B}}{L}\int^{L/2}_{-L/2}\int^{\infty}_{0} dx dE \mathcal{N}(x,E,T,\phi)\\
\times\left\{f(E)\ln[f(E)]+[1-f(E)]\ln[1-f(E)]\right\},\nonumber
\end{eqnarray}
and $f(E)=\left\{1+\exp[E/(k_{\text{B}}T)]\right\}^{-1}$ is the Fermi-Dirac quasiparticle distribution function at temperature $T$. 
We introduce the position-dependent specific heat ($c$) and entropy ($s$) according to the expressions
\begin{eqnarray}
C(T,\phi)&=&\frac{1}{L}\int^{L/2}_{-L/2}dx \,\,c(x,T,\phi),\nonumber\\
{\cal S}(T,\phi)&=&\frac{1}{L}\int^{L/2}_{-L/2}dx \,\,s(x,T,\phi).
\label{xCS}
\end{eqnarray}

In order to allow  a comparison of our results with the experiments it is important to calculate the total specific heat of the loop.
To this end  one should also add the electronic contribution of the superconducting part of the ring (which can be rather small due to the exponential suppression with temperature), as well as the contributions from the lattice phonons of the N and S parts. 
We will discuss this issue at the end of the paper.

\section{Results}
\label{res}
In this section we report our numerical results for entropy and specific heat determined by using Eqs.~(\ref{Cv}) and (\ref{Sss}) and the expression for the DOS.
The latter is calculated by numerically solving the Usadel equations, using the relaxation method~\cite{Press}, in one of the above parametrizations.

In the following analysis we shall focus on \emph{long} SNS junctions only ($\Delta_0\gg E_{Th}$), as this is the relevant regime for metallic diffusive Josephson junctions. 
\begin{figure}[t]
\includegraphics[width=\columnwidth,clip]{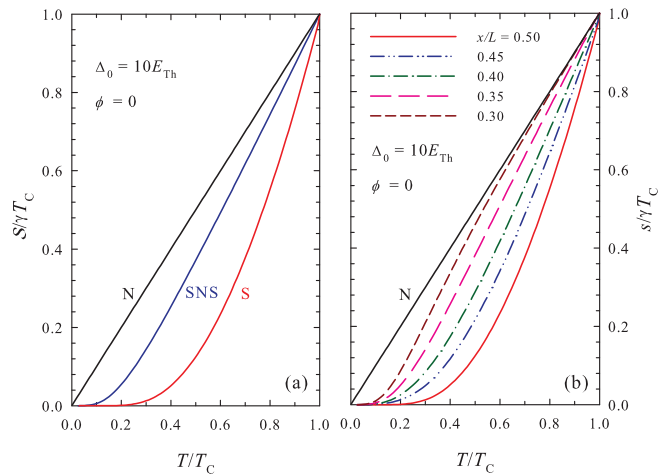}	\caption{(Color online) (a) Entropy ${\cal S}$ vs temperature $T$ compared to that of a normal metal (N) and of a superconductor (S) calculated for $\Delta_0=10E_{{\rm Th}}$ and $\phi=0$.  (b) Local entropy $s$ vs $T$  at different positions  $x$ along the N region calculated for $\Delta_0=10E_{{\rm Th}}$ and $\phi=0$. We have set $r=0$, $\Gamma_{\rm in}=0$, $\Gamma_{\rm sf}=0$ and $\Gamma_{\rm S}=0$, while ${\cal S}$ and $s$ are normalized with respect to the N metal value.
$T_{\text{C}}$ is the critical temperature of the superconductor.}
\label{fig2}
\end{figure}
The main general feature is that the DOS is an even function of energy, exhibiting a minigap $E_g$~\cite{zhou}, and it is position-dependent.
Moreover, $E_g$ depends on Thouless 
energy,  phase difference, inelastic and spin-flip scattering rate, and ratio $r$~\cite{Hammer}.
In particular, $E_g$ is maximum at $\phi=0$, where $E_g\approx 3.12 E_{\text{Th}}$ in the limit of long SNS junctions, whereas it vanishes  at $\phi=\pi$~\cite{zhou}.
As we shall see, entropy, and therefore specific heat, are heavily affected by the magnitude of the minigap,
which is the most evident manifestation of the proximity effect. 

The electronic entropy of the proximized N region of the SNS junction versus temperature is compared with that of a normal metal (N) and of a superconductor (S) in Fig.~\ref{fig2}(a).
Here we set $\Delta_0=10E_{\text{Th}}$ and $\phi=0$ and normalize $\mathcal{S}$ to the N case value, namely $\mathcal{S}_{\text{N}}=\gamma T$, where $\gamma=2 \pi^2{\cal N}_{\rm F}k_{\rm B}^2/3$.
For the case of a superconductor, the figure shows the standard BCS result~\cite{Tinkham}, where the electronic entropy is exponentially damped for $T\ll T_{\text{C}}$, stemming from the presence of the gap $\Delta$ in the DOS.
Figure~\ref{fig2}(a) shows that in the proximized region $\mathcal{S}$ strongly departs from the normal-metal behaviour especially at low temperature. In particular, for $T\ll T_{\text{C}}$, $\mathcal{S}$ exhibits a strong suppression that resembles that one of a superconductor. It turns out that the temperature determining the onset of such suppression is set by the proximity-induced minigap $E_g$. 
More precisely, for $k_{\rm B}T\ll E_{\text{Th}}$ and in the long junction regime ($\Delta_0\gg E_{\text{Th}}$), for $\phi=0$ the electronic entropy can be described rather well by the following expression
\begin{equation} 
\mathcal{S}(T,\phi=0)=\mathcal{S}_{\text{N}}\text{exp}(-aE_{\text{Th}}/k_{\text{B}}T),
\end{equation}
where for the prefactor we get $a\approx2.1$. Accordingly, in the same limits we get the following approximated expression for the electronic specific heat:
\begin{equation}
C(T,\phi=0)\simeq C_{\text{N}}(aE_{\text{Th}}/k_{\text{B}}T)\text{exp}(-aE_{\text{Th}}/k_{\text{B}}T),
\end{equation}
where $C_{\text{N}}=\gamma T$ is the specific heat in the normal state.

In Fig.~\ref{fig2}(b) the position-dependent electronic entropy $s(x,T,\phi)$ is plotted as a function of temperature for different values of $x$.
First of all we note that at the NS interface ($x=L/2$)
$s$ coincides with that of the superconducting lead. Secondly, by moving toward the centre of the N region, $s$ tends to approach the entropy of a N metal, while the temperature which determines the onset of strong suppression gets reduced. This latter behaviour stems from the weakening of the  proximity effect (i.e., reduction of superconducting correlations) departing from the NS interfaces.

\begin{figure}[t]
\includegraphics[width=\columnwidth,clip]{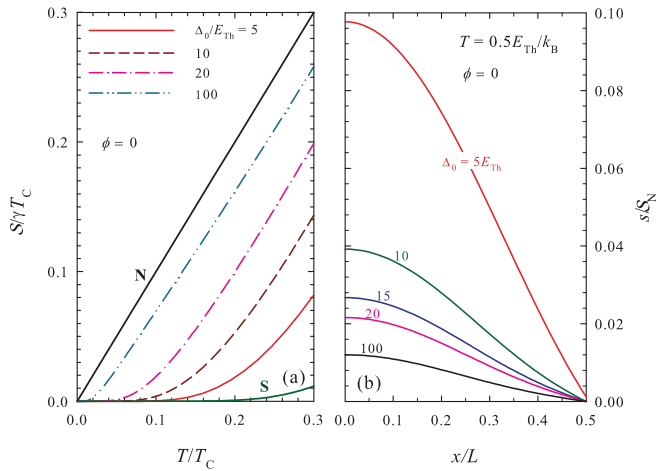}
\caption{(Color online) (a) Entropy ${\cal S}$ vs $T$  at  $\phi=0$ calculated for several ratios $\Delta_0/E_{\text{Th}}$. (b) $s$ vs $x$ at $\phi=0$ and $T=0.5 E_{{\rm Th}}/k_{\text{B}}$ calculated for several ratios $\Delta_0/E_{\text{Th}}$. We have set $r=0$, $\Gamma_{\rm in}=0$, $\Gamma_{\rm sf}=0$ and $\Gamma_{\rm S}=0$, while ${\cal S}$ and $s$ are normalized as in Fig.~\ref{fig2}. N and S curves are reported for comparison.}
\label{fig3}
\end{figure} 
\par
Figure~\ref{fig3}(a) shows the value of ${\cal S}$ versus $T$ calculated for several lengths of the N region expressed by the ratio $\Delta_0/E_{\rm Th}$ at $\phi=0$.
As expected, for long junctions ($\Delta_0=100 E_{\rm Th}$) the curve approaches the N metal entropy, which follows from weakening of the proximity effect in longer SNS junctions. 
Similarly, also the temperature which sets the onset of strong suppression gets reduced for longer junctions, as the minigap in the DOS gets suppressed by increasing $L$.
In the opposite limit ($\Delta_0=5 E_{\rm Th}$), the entropy approaches that in the S state so that shorter SNS junctions are more suitable in order to maximize the effect of proximization on the entropy.  

The behaviour of the position-dependent electronic
entropy $s$ for fixed length $L$ allows to investigate the role of different superconducting gaps, and it is shown in Fig.~\ref{fig3}(b). 
Here we set $T=0.5E_{\rm Th}/k_{\text{B}}$ and $\phi=0$. 
The entropy strongly decreases by moving away from the middle of the junction $(x=0)$, approaching the value in the superconductor at the NS interface ($x=L/2$).
We emphasize that the increase of the value of $\Delta_0$ leads to an overall strong suppression of the entropy. 
This can be understood by noting that, for a given $L$, smaller-gap superconductors reduce the proximity effect in the N region producing an enhancement of the electronic entropy. 

\begin{figure}[t]
\includegraphics[width=\columnwidth,clip]{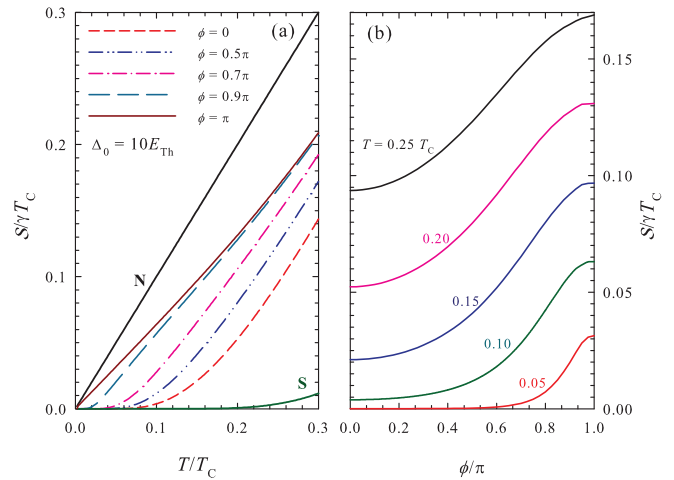}
\caption{(Color online) (a) Entropy ${\cal S}$ vs $T$ calculated for $\Delta_0=10E_{{\rm Th}}$ and different values of $\phi$ .  (b) Entropy ${\cal S}$ vs $\phi$ calculated for $\Delta_0=10E_{{\rm Th}}$ and several values of temperature. We have set $r=0$, $\Gamma_{\rm in}=0$, $\Gamma_{\rm sf}=0$ and $\Gamma_{\rm S}=0$, while ${\cal S}$ is normalized as in Fig.~\ref{fig2}. N and S curves are reported for comparison.}
\label{fig4}
\end{figure} 

Figure~\ref{fig4}(a) displays the electronic entropy ${\cal S}$ as a function of $T$ for $\Delta_0=10E_{\rm Th}$ and several different values of $\phi$. The curves for $\phi=\pi$ were obtained through the Riccati parametrization which allows stable solutions of the Usadel equations.
The data show a strong phase-dependence of ${\cal S}$ stemming from the coherent nature of proximity effect.
In particular, by increasing $\phi$ from zero one finds an entropy enhancement in the whole temperature range mainly due to the suppression of the minigap $E_g$. 
Remarkably, at $\phi=\pi$ the minigap vanishes and, for temperatures below $T = 0.15 T_{\rm C}$, the entropy shows a linear behaviour, namely  ${\cal S}/\gamma \cong 0.65T$, which corresponds to a $35\%$ reduction with respect to the entropy of a N metal ${\cal S}_{\rm N}$. 
The full phase dependence is shown in Fig.~\ref{fig4}(b), where ${\cal S}$ versus phase is plotted for $\Delta_0=10E_{\rm Th}$ at several different temperatures. In particular, for each temperature the electronic entropy increases by increasing the phase difference between the superconductors,  and it turns out to be maximized at $\phi=\pi$, i.e., when the minigap is zero. We stress that the overall entropy variation with $\phi$, especially at the lowest temperatures, can be very large.
This directly reflects on the electronic specific heat phase dependence.  

\begin{figure}[t]
\includegraphics[width=\columnwidth,clip]{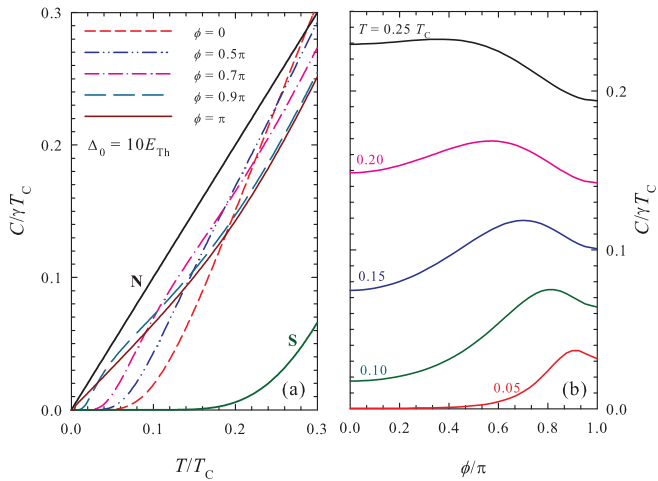}
\caption{(Color online) (a) Electronic specific heat $C$ vs $T$ calculated for  $\Delta_0=10E_{{\rm Th}}$ and different values of $\phi$.  (b) $C$ vs $\phi$ calculated for $\Delta_0=10E_{{\rm Th}}$ and several values of temperature. We have set $r=0$, $\Gamma_{\rm in}=0$, $\Gamma_{\rm sf}=0$ and $\Gamma_{\rm S}=0$, while $C$ is normalized with respect to the N metal value. N and S curves are reported for comparison.}
\label{fig5}
\end{figure}

Figure~\ref{fig5}(a) shows the electronic specific heat $C$ vs temperature for a junction with $\Delta_0=10E_{\text{Th}}$ at different values of the phase difference between the superconductors.
In this calculations we extended the results of Ref.~\onlinecite{Rabani} up to $\phi=\pi$.
A strong phase-dependence is found at all temperatures: by increasing $\phi$ from zero to $\pi$, $C$ tends to increase for small temperatures and to decrease for temperatures close to $T_{\rm C}$.
A linear temperature dependence is found for $\phi=\pi$ below $T = 0.15 T_{\rm C}$.

In the next subsections we will discuss the influence of inelastic and spin-flip scattering, the presence 
of a barrier at the NS contacts, and the presence of subgap states within the gap induced by a finite value of $\Gamma_{\rm S}$.
For the sake of definiteness, in the following we will focus on the case 
$\phi=0$ and $\Delta_0=10E_{\rm Th}$, which shows the largest deviation from the N metal case.

\subsection{Inelastic scattering}

\begin{figure}[t]
\includegraphics[width=\columnwidth,clip]{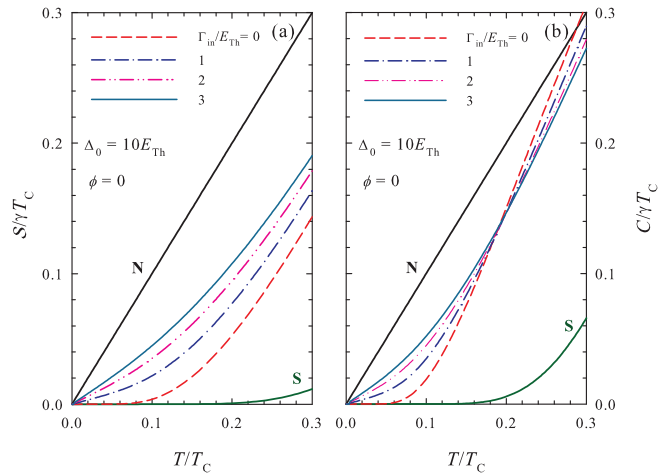}
\caption{(Color online) Influence of inelastic scattering on (a) entropy ${\cal S}$ and (b) specific heat $C$ for  $\Delta_0=10 E_{\rm Th}$, $\phi=0$ and several values of $\Gamma_{\rm in}$. We have set $r=0$, $\Gamma_{\rm sf}=0$ and $\Gamma_{\rm S}=0$, while ${\cal S}$ is normalized as in Fig.~\ref{fig2} and $C$ as in Fig.~\ref{fig5}. N and S curves are reported for comparison.}
\label{fig6}
\end{figure}

Inelastic scattering of quasiparticles in the N region is introduced through the parameter $\Gamma_{\rm in}$ appearing in the Usadel equation. Here we assume that, at low temperatures, the inelastic scattering is independent on temperature.
A non-zero value of $\Gamma_{\rm in}$ affects the thermodynamic quantities by inducing a finite DOS below the mingap.
Figures~\ref{fig6}(a) and \ref{fig6}(b) show the entropy and specific heat,  respectively, as a function of temperature for several different values of $\Gamma_{\rm in}$. 
The curves corresponding to the normal and superconducting states are displayed for comparison in the same plots.
As expected, for increasing inelastic scattering the entropy increases at all temperatures, while the specific heat increases for $T\lesssim 0.2 T_{\rm C}$ but decreases for larger temperatures.
Note, however, that at low temperatures $\Gamma_{\rm in}$ is typically~\cite{Belzig2} of the order of $10^{-3}\Delta$, which corresponds to $\Gamma_{\rm in}\sim 0.01E_{\rm Th}$ when $\Delta_0=10E_{\rm Th}$.
In turns out that such small value has negligible effect on the entropy and specific heat of the system.

\subsection{Spin-flip scattering}
\begin{figure}[b]
\includegraphics[width=\columnwidth,clip]{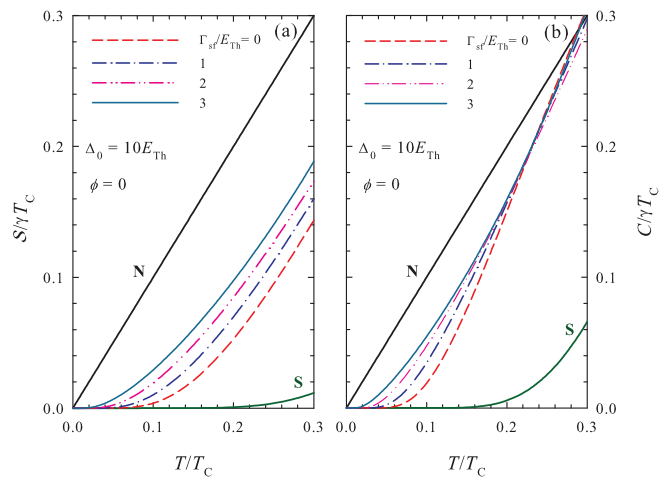}
\caption{(Color online) Influence of spin-flip scattering on (a) entropy ${\cal S}$ and (b) specific heat $C$ for $\Delta_0=10 E_{\rm Th}$, $\phi=0$ and several values of $\Gamma_{\rm sf}$. We have set $r=0$, $\Gamma_{\rm in}=0$ and $\Gamma_{\rm S}=0$, while ${\cal S}$ is normalized as in Fig.~\ref{fig2} and $C$ as in Fig.~\ref{fig5}. N and S curves are reported for comparison.}
\label{fig7}
\end{figure}
We consider here the effect of pair-breaking mechanisms by setting a finite spin-flip rate $\Gamma_{\rm sf}$ in the normal metal, while assuming that $\Gamma_{\rm sf}=0$ in the superconductors.
According to Ref.~\onlinecite{Hammer}, in the presence of spin-flip scattering the DOS is suppressed and the minigap gets reduced, disappearing eventually.
One should, therefore, expect a change in the entropy and specific heat.
Figures~\ref{fig7}(a) and (b) show the plots of ${\cal S}$ and $C$, respectively, for various values of $\Gamma_{\rm sf}$.
Indeed, by increasing $\Gamma_{\rm sf}$, the entropy increases for all temperatures while the specific heat increases only in a smaller temperature range.
More precisely, at $T\approx 0.21 T_{\rm C}$ the spin-flip scattering does not modify the 
specific heat as all plots cross each other in this point [see Fig. \ref{fig7}(b)].
We remark that, as stated in Sec.~\ref{ricc}, the spin-flip rate is related to the intensity of the magnetic field applied perpendicularly to the loop and needed to establish the phase bias.
Using the numerical values given above we find that $\Gamma_{\rm sf}$ equals $E_{\rm Th}$ only for a large magnetic field.
Namely, $\Gamma_{\rm sf}=E_{\rm Th}$ for $H=281 {\rm G}=2.6 H_C$, where $H_C=105 $G is the aluminium critical magnetic field. 
For  small magnetic field no relevant changes must be expected.

\subsection{Non-ideal NS interfaces}
\begin{figure}[t]
\includegraphics[width=\columnwidth,clip]{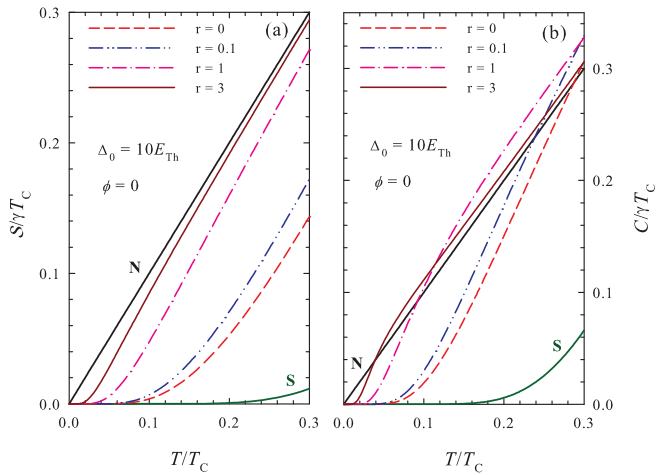}
\caption{(Color online) Influence of non-ideal interfaces on (a) entropy ${\cal S}$ and (b) specific heat $C$ for $\Delta_0=10 E_{\rm Th}$, $\phi=0$ and several values of $r$. We have set $\Gamma_{\rm in}=0$, $\Gamma_{\rm sf}=0$ and $\Gamma_{\rm S}=0$, while ${\cal S}$ is normalized as in Fig.~\ref{fig2} and $C$ as in Fig.~\ref{fig5}. N and S curves are reported for comparison.}
\label{fig8}
\end{figure}

The presence of a barrier at the NS boundaries is taken into account through the ratio of resistances $r$ and the transmission probability of the barrier $\tau$. Their effect, on both DOS and supercurrent, in a SNS structure has been considered in Ref.~\onlinecite{Hammer}.
It was found that, on the one hand, by increasing $r$ from zero not only the DOS but also the minigap gets suppressed. On the other, the minigap only slightly depends on the value of $\tau$.
This motivates us to explore the effect of a finite $r$ on entropy and specific heat, while keeping $\tau=1$.
In Figs.~\ref{fig8}(a) and \ref{fig8}(b) we report the plots of entropy and specific heat, respectively, for various values of the ratio $r$.
It can be noted that both ${\cal S}$ and $C$ are strongly affected by a finite value of $r$. Indeed, already for $r=3$ the curves are quite close to the one relative to N apart from very small temperatures.
We conclude that very small values of the ratio $r$ are necessary to observe a clear phase-dependent specific heat.

\subsection{Subgap states}
\begin{figure}[t]
\includegraphics[width=\columnwidth,clip]{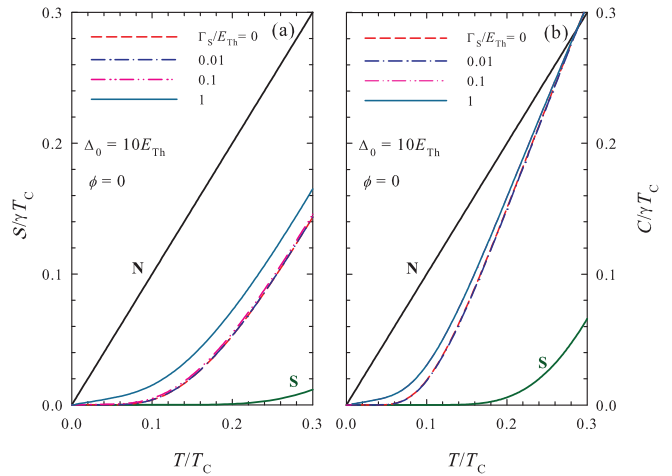}
\caption{(Color online) Influence of quasiparticle subgap states on (a) entropy ${\cal S}$ and (b) specific heat $C$ for $\Delta_0=10 E_{\rm Th}$, $\phi=0$ and several values of $\Gamma_{\rm S}=0$. We have set $\Gamma_{\rm in}=0$, $\Gamma_{\rm sf}=0$ and $r=0$, while ${\cal S}$ is normalized as in Fig.~\ref{fig2} and $C$ as in Fig.~\ref{fig5}. N and S curves are reported for comparison.}
\label{fig9}
\end{figure}

The effect of the presence of quasiparticles states within the gap can be studied by setting $\Gamma_{\rm S}\neq0$ 
in Eq.~(\ref{bc}).
Figures~\ref{fig9}(a) and \ref{fig9}(b) show the plots of entropy and specific heat, respectively, as a function of temperature for various values of $\Gamma_{\rm S}$. 
The effect of a finite $\Gamma_{\rm S}$ is very weak up to $\Gamma_{\rm S}\approx0.1E_{\rm Th}$.
Since the typical experimental values are between $\Gamma_{\rm S}=10^{-3}\Delta_0$ and $\Gamma_{\rm S}=10^{-5}\Delta_0$~\cite{pekola},
these results demonstrate that virtually no influence can be expected from quasiparticles states within the gap.

\subsection{Contribution of the superconducting loop}
\begin{figure}[t]
\includegraphics[width=\columnwidth,clip]{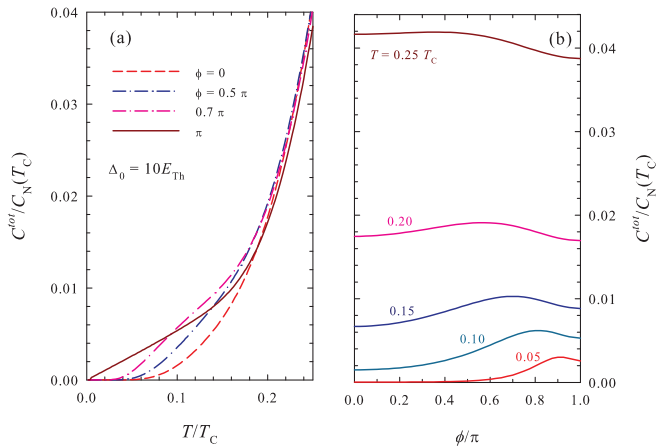}
\caption{(Color online) Total specific heat of the loop $C^{\rm tot}$, calculated including the contribution from the S regions and the phonons, as a function of (a) temperature and (b) phase-difference. $C^{\rm tot}$ is normalized with respect of the normal-state total specific heat evaluated at the critical temperature of Al. Plots are relative to the following parameters: $\Delta_0=10 E_{\rm Th}$ and $r=0$.}
\label{fig10}
\end{figure}

Electronic specific heat can be measured by ac calorimetry as discussed in Ref.~\onlinecite{Fominaya}. 
Recent developments in highly-sensitive calorimetry applied to mesoscopic systems allow to measure 
small signals \cite{Bourgeois,Ong}.
In this section we estimate the specific heat which would be measured in an actual experimental setup, including the phonons contribution, aimed at observing the tuning of the specific heat by changing the phase difference between the S electrodes.
To determine the total specific heat we suppose that the system consists of a SQUID loop of square 
shape (i.e., like the one depicted in Fig.~\ref{fig1}) with side length equal to $1\mu{\rm m}$, N region length $L=0.6 \mu{\rm m}$, width $W=0.1 \mu{\rm m}$ and constant thickness.
We assume that the normal material is Cu while the superconductor is Al, and that the superconducting coherence length $\xi\simeq 190$ nm.
The total specific heat $C^{\rm tot}$ can be calculated through the expression
\begin{equation}\label{aver}
C^{\rm tot} V^{\rm tot}=(C^{\rm Cu}+\kappa^{\rm Cu}T^3)V^{\rm Cu}+(C^{\rm Al}+\kappa^{\rm Al}T^3)V^{\rm Al},
\end{equation} 
where the second terms in the round brackets account for the low-temperature phononic contribution.
Here
\begin{equation}
\kappa^{\rm Cu (Al)}=\frac{2\pi^2}{5} \frac{k_{\rm B}^4}{(\hbar v^{\rm Cu (Al)})^3},
\end{equation} 
$v^{\rm Cu (Al)}$ being the sound velocity in Cu (Al)~\cite{Ashcroft}.
$V^{\rm Cu}$ and $V^{\rm Al}$ are the volumes relative to the N region (Cu) and the S region (Al), respectively.
$V^{\rm tot}$ is the total volume of the structure.
In Figs.~\ref{fig10} we collect the results for the total specific heat normalized to its normal state value calculated at Al critical temperature $T_{\rm C}$: $C_{\rm N}^{\rm tot} V^{\rm tot}=(\gamma^{\rm Cu}T_{\rm C}+\kappa^{\rm Cu}T_{\rm C}^3)V^{\rm Cu}+(\gamma^{\rm Al}T_{\rm C}+\kappa^{\rm Al}T_{\rm C}^3)V^{\rm Al}$ for $\Delta_0=10 E_{\rm Th}$.
As shown in Fig.~\ref{fig10}(a), where plots of $C^{\rm tot}$ as a function of temperature are reported for various values of $\phi$, the overall specific heat maintains a different functional dependence on $T$ for the different values of phase-difference.
In Fig.~\ref{fig10}(b) $C^{\rm tot}$ is plotted as a function of $\phi$ for various temperatures.
The comparison between Fig.~\ref{fig10}(b) and Fig.~\ref{fig5}(b) makes clear that the total specific heat is still strongly phase-dependent, despite the inclusion of the contributions from the S regions and the phonons. 
We conclude that phase-dependent specific heat can be measured, and that can be maximized in the case of  N regions of intermediate length ($\Delta_0=10 E_{\rm Th}$).
We emphasize that while in the case of \emph{very} long junctions ($\Delta_0\gg E_{\rm Th}$) the specific heat is close to the normal state value and weakly phase-dependent, for \emph{short} junctions ($\Delta_0\ll E_{\rm Th}$) the specific heat of the S regions dominates therefore masking the dependence on the phase.

\section{Conclusion}
\label{conc}
In summary, in an ideal SNS mesoscopic junction both the electronic entropy and the specific heat exhibit a large deviation with respect to the normal and superconductor case, and show a marked phase-dependence~\cite{Rabani}.
In this paper we have studied in details such thermodynamic quantities in a realistic SNS junction as a function of temperature and superconducting phase difference $\phi$, aiming at investigating the role of non-idealities occurring in an actual experiment.
Indeed, the presence of barriers at the NS contacts, as well as finite inelastic and spin-flip scattering produce a weakening of the proximity effect which is expected to affect both entropy and specific heat and to suppress their phase-dependence.

Here we have focused on the intermediate junction length, with $\Delta_0=10E_{\rm Th}$, which turns out to be a favorable condition as far as phase-dependence visibility is concerned.
As a general feature, we have shown that, at low temperatures, the entropy presents an exponential suppression which resembles the one of a superconductor and where the role of the superconducting gap is played by the minigap.
The most important results are the following.
First, we have found that inelastic scattering in the N region negligibly affect the specific heat, at least at the typical experimental values relative to low temperatures. We arrive at the same conclusion also for subgap quasiparticle states in the superconductors.
Similarly, the presence of spin-flip scattering, related to the applied perpendicular magnetic field, only slightly changes the entropy and specific heat of the system.
Second, we have found that the effect of barriers at the NS interfaces is strong and small values of barrier resistance, compared to the resistance of the N region, are necessary to observe a clear phase-dependent specific heat.
Furthermore, assuming that the measurement is performed employing the method reported in Refs.~\cite{Bourgeois,Ong}, we have taken into account the additional contributions to the measured specific heat in the actual experimental setup, in which the SNS junction is embedded in a superconducting loop. Using Al as superconductor and Cu as normal-metal  material, with reasonable geometry and sizes, we checked that a clear phase-dependent signal is measurable. 
We note finally that our results for the electronic specific heat can be relevant for phase tuning the thermodynamic properties of mesoscopic devices like, for instance, superconducting single-photon sensors based on proximity effect \cite{PJS}.

\begin{acknowledgments}
We thank O. Bourgeois for his contribution at the early stage of this work.
Furthermore, we acknowledge C. J. Cuevas and M. A. Shahzamanian for fruitful discussions, the Iranian Nanotechnology Initiative and the NanoSciERA "NanoFridge" project of the EU for partial financial support.
\end{acknowledgments}


\begin{thebibliography}{99}
\bibitem{degennes}
				P. G. de Gennes and D. Saint-James, Phys. Lett. \textbf{4}, 151 (1963).
\bibitem{editorHekking}
	{\it Mesoscopic Superconductivity}, Proceedings of the NATO ARW,
	F.\,W.\,J. Hekking, G. Sch\"on, and D.\,V. Averin, eds.,
	Physica B {\bf 203} (1994).
\bibitem{Beenakker95}
	C. W. J. Beenakker, Rev. Mod. Phys. {\bf 69}, 731 (1997).
\bibitem{Lambert96}
	C. J. Lambert and R. Raimondi, J. Phys. Cond. Matt. {\bf 10}, 901 (1998).
\bibitem{Belzig} 
        W. Belzig, F. K. Wilhelm, C. Bruder, G. Sch\"on, and A. D. Zaikin, 
        Superlatt. Microstruct. {\bf 25}, 1251 (1999).
\bibitem{Taddei} 
        F. Taddei, F. Giazotto, and R. Fazio, J. Comput. Theor. Nanosci. 
        \textbf{2}, 329 (2005).
\bibitem{ds}
				F. Giazotto, P. Pingue, F. Beltram, M. Lazzarino, D. Orani, S. Rubini, and A. Franciosi, Phys. Rev. Lett. \textbf{87}, 216808 (2001). 

\bibitem{Giazotto} 
        F. Giazotto, T. T. Heikkil\"a, A. Luukanen,  A. M. Savin and J. P. Pekola, 
        Rev. Mod. Phys. \textbf{78}, 217 (2006).
\bibitem{Rabani} 
        H. Rabani, F. Taddei, O. Bourgeois, R. Fazio and F. Giazotto, Phys. Rev. B, {\bf 78}, 012503 (2008).
\bibitem{Bourgeois}  
        O. Bourgeois S. E. Skipetrov F. Ong and J. Chaussy, 
        Phys. Rev. Lett. {\bf 94}, 057007 (2005).
\bibitem{Ong} 
        F. R. Ong and O. Bourgeois, Europhys. Lett. {\bf 79}, 67003 (2007).
\bibitem{Usadel}
		K. D. Usadel, Phys. Rev. Lett. {\bf 25}, 507 (1970).
\bibitem{Laakso}  
        M. A. Laakso, P. Virtanen, F. Giazotto, and T. T. Heikkil\"a, Phys. Rev. B \textbf{75}, 094507 (2007).
\bibitem{pekola}
	J. P. Pekola, T. T. Heikkil\"a, A. M. Savin, J. T. Flyktman, F. Giazotto, and F. W. J. Hekking,
	Phys. Rev. Lett. {\bf 92}, 056804 (2004).
\bibitem{Kupriyanov}
		M. Yu. Kupriyanov, and V. F. Lukichev, Zh. Eksp. Teor. Fiz. {\bf 94}, 139 (1988) [Sov. Phys. JETP 	{\bf 67}, 1163 (1988)].
\bibitem{Schopohl} 
        N. Schopohl and K. Maki, Phys. Rev. B \textbf{52}, 490 (1995).
\bibitem{Eschrig} 
        M. Eschrig, Phys. Rev. B \textbf{61}, 9061 (2000).
\bibitem{Hammer} 
        J. C. Hammer, J. C. Cuevas, F. S. Bergeret, and W. Belzig, Phys. Rev. B \textbf{76}, 064514 (2007).
\bibitem{Belzig2} 
        W. Belzig, C. Bruder and G. Sch\"on, Phys. Rev. B {\bf 54}, 9443 (1996).
\bibitem{Press} 
        W. H. Press, S. A. Teukolsky and W. T. Vetterling, \emph{Numerical Recipes in Fortran}, 
        2nd ed. (Cambridge University  Press, 1992).
\bibitem{zhou}
	F. Zhou, P. Charlat, B. Spivak, and B. Pannetier, J. Low Temp. Phys. \textbf{110}, 841 (1998).
\bibitem{Tinkham} 
        M. Tinkham, {\em Introduction to Superconductivity}, 2nd ed. (McGraw-Hill, Inc., Singapore, 1996). 
\bibitem{Ashcroft}
	N. W. Ashcroft and N. D. Mermin,  {\em Solid State Physics}, (Saunders College Publishing, Orlando, 1976).
\bibitem{Fominaya} 
        F. Fominaya , T. Fournier, P. Gandit and J. Chaussy, Rev. Sci. Instrum. \textbf{68}, 4191 (1997).

\bibitem{PJS}
				F. Giazotto, T. T. Heikkil\"a, G. P. Pepe, P. Helist\"o, A. Luukanen, and J. P. Pekola, Appl. Phys. Lett. \textbf{92}, 162507 (2008).



\end{thebibliography}
\end{document}